\documentclass[11pt]{article}
\pdfoutput=1
\usepackage{epsfig,amscd,amssymb,hyperref,titling}
\usepackage{amsmath,graphicx,comment,mathtools}

\usepackage[dvipsnames]{xcolor}

\usepackage[normalem]{ulem}
\usepackage[sort&compress, numbers]{natbib}
\usepackage{tikz}

\numberwithin{equation}{section}

\usepackage{tikz}
\usetikzlibrary{calc}
\usetikzlibrary{decorations.markings}
\usetikzlibrary{decorations.pathmorphing, arrows}
\usetikzlibrary{arrows.meta}
\newcommand{\Sascha}[1]{{\color{red}\ifmmode\text{\footnotesize(FE) #1}\else\footnotesize{(SG) #1}\fi}}
\def\df{{f}}
\newcommand{\paul}[1]{{\color{blue}\ifmmode\text{\footnotesize(PF) #1}\else\footnotesize{(PF) #1}\fi}}

\begin{document}
\begin{center}
    \Large{\bf XYZ integrability the easy way}
\end{center}
	\date{\today}
	\begin{center}
	Paul Fendley$^{1,2}$, Sascha Gehrmann$^1$, Eric Vernier$^3$ and Frank Verstraete$^{4,5}$
\end{center}	
\begin{center}
{\small
{\bf 1}
The Rudolf Peierls Centre for Theoretical Physics, University of Oxford, Oxford OX1 3PU, UK\\
{\bf 2} All Souls College, University of Oxford\\
{\bf 3}  Université Paris Cité and Sorbonne Université, CNRS, Laboratoire de
  Probabilités, Statistique et Modélisation, F-75013 Paris, France\\
{\bf 4} DAMTP, University of Cambridge, Wilberforce Road,
Cambridge, CB3 0WA, UK\\
{\bf 5} Department of Physics and Astronomy, Ghent University,  Belgium}
\end{center}
\section*{Abstract}

Sutherland showed that the XYZ quantum spin-chain Hamiltonian commutes with the eight-vertex model transfer matrix, so that Baxter's subsequent {\em tour de force} proves the integrability of both. The proof requires parametrising the Boltzmann weights using elliptic theta functions and showing they satisfy the Yang-Baxter equation. We here give a simpler derivation of the integrability of the XYZ chain by explicitly constructing an extensive sequence of conserved charges from a matrix-product operator. We show that they commute with the XYZ Hamiltonian with periodic boundary conditions or an arbitrary boundary magnetic field. A straightforward generalisation yields impurity interactions that preserve the integrability. Placing such an impurity at the edge gives an integrable generalisation of the Kondo problem with a gapped bulk. We make contact with the traditional approach by relating our matrix-product operator to products of the eight-vertex model transfer matrix. 
	
\date{\today}
%\tableofcontents
\bigskip

%%%%%%%%%%%%%%%%%%%%%%%
\section{Introduction}
%%%%%%%%%%%%%%%%%%%%%%%

Bethe's application of his famous ansatz \cite{Bethe1931} showed that the Heisenberg quantum spin-$\tfrac12$ chain is integrable. Quite a time passed before this landmark in theoretical physics was extended. The XXZ chain is the most obvious generalisation of Heisenberg's, with couplings deformed to break the SU(2) symmetry to U(1)\,$\ltimes$\,$\mathbb{Z}_2$. The extension of the Bethe-ansatz calculation to this case is relatively straightforward \cite{Orbach1958} because the U(1) symmetry yields a simple ``reference state", i.e.\ an exact eigenstate from which the eigenstates are built by acting with plane-wave operators. In the '60s, many important extensions of the Bethe ansatz to both classical and quantum models were found, but in all such cases at least a U(1) symmetry was present.

In the XYZ quantum spin chain, the symmetry is broken further.
%The spin-$\tfrac12$ XYZ chain is comprised of $L$ two-state quantum systems. 
Its Hamiltonian is built from operators $\sigma^a_j=1\otimes 1 \cdots \otimes\sigma^a \otimes\cdots \otimes1$ with the Pauli matrix $\sigma^a$ acting non-trivially on the two-state system at site $j$. With free boundary conditions at both ends of an $L$-site chain, it is
\begin{align}
H_{\rm free}= \sum_{j=1}^{L-1} h_{j,j+1}\ ,\qquad\ h_{j,j+1}=J_x\, \sigma^x_j\sigma^x_{j+1}+J_y\, \sigma^y_j\sigma^y_{j+1}
+ J_z\,\sigma^z_j\sigma^z_{j+1}\ .
\label{HXYZ}
\end{align}
The XXZ case corresponds to equating the couplings $J_x$\,=\,$J_y$, with all three the same for the Heisenberg chain. The reference state for the Bethe ansatz then can be taken to be one of the fully polarised states. However, for arbitrary couplings the symmetry is reduced to two $\mathbb{Z}_2$ symmetries, generated by the spin-flip operator $\prod_j \sigma^x_j$ along along with the $z$-parity $\prod_j \sigma^z_j$. The XYZ chain in general does not admit an obvious Bethe ansatz, as the fully polarised states are no longer exact eigenstates. 

Nonetheless, via a still-stunning {\em tour de force}, Baxter invented several methods to find the Bethe equations for the eigenstates of the eight-vertex model, a classical two-dimensional spin model with the same symmetries \cite{Baxter1973a,Baxter1973b,Baxter1973c}.
In doing so he introduced the Yang-Baxter equation in its modern form, where the Boltzmann weights depend on the ``spectral" parameter. Weights solving this equation yield transfer matrices that commute at different values of this parameter. This equation has played a central role in studies of integrability ever since, with ramifications far from its original setting.

Quantum spin-chain Hamiltonians fit naturally into the Yang-Baxter story.  Typically (but not always) one can tune the spectral parameter to yield a trivial transfer matrix, with the first correction yielding a local quantum Hamiltonian. Indeed, prior to Baxter's work, Sutherland had shown that the periodic XYZ Hamiltonian commutes with the eight-vertex model \cite{Sutherland1970}. The integrability of the latter thus immediately implies that of the former. This intimate connection means that analysis of the two can be done in tandem. From this point of view, the 2d classical model is more fundamental, as it contains the spectral parameter. The quantum spin chain arises straightforwardly as a limit.

The purpose of this paper is to demonstrate the integrability of the XYZ chain in a simple and direct fashion, without recourse to any 2d local classical model. We find a {\em matrix product operator} (MPO) \cite{Verstraete2004,Pirvu_2010,Cirac2020} commuting with the XYZ Hamiltonian. Its construction is elementary, and applies both to periodic boundary conditions and to open ones with an arbitrary boundary magnetic field. This MPO depends on a free parameter akin to the spectral parameter, and expanding it in this parameter yields a series of quantities commuting with the Hamiltonian, extensive in system size. %It is easily verified that these quantities are not simply powers of the Hamiltonian, and so require the integrability of the XYZ Hamiltonian. 

The key to finding this MPO comes from a seemingly disconnected result.  Systems with topological order often have degenerate ground states not related by any obvious symmetry. Nonetheless, one typically can construct ``zero mode'' operators mapping between them. Moreover, certain gapped systems have an even stronger characteristic: they possess {\em strong zero mode} operators that commute with the Hamiltonian up to exponentially small finite-size corrections \cite{Alicea2015,Fendley2015}. When these operators do not commute with a 
global symmetry, they result in degeneracies between different symmetry sectors. The simplest example of such 
an operator occurs in the Ising model, where it follows easily from the free-fermion solution \cite{Pfeuty1970,Kitaev2000}. A much less obvious example of a strong zero mode occurs in the XYZ spin chain, where it was originally found by brute force \cite{Fendley2015}.

The connection of this strong zero mode operator to the integrability of the XYZ chain looks rather mysterious in its original formulation as a power series localised by one of the edges. Nonetheless, we show in this paper how a straightforward generalisation of this operator yields a generating function for an extensive sequence of conserved charges commuting with the XYZ Hamiltonian. Namely, we rewrite it in terms of what turns out to be a rather elegant MPO. This expression suggests a natural generalisation, which we show is a generating function for an extensive hierarchy of conserved quantities in the XYZ chain. 

Such conserved currents characterise an integrable Hamiltonian. Our construction thus provides a direct proof of the integrability of the XYZ chain, without recourse to the eight-vertex model. A feature is that we never need to write the Hamiltonian in terms of elliptic theta functions, making manipulations easier. We straightforwardly find an integrable impurity Hamiltonian with our approach, and putting the impurity at the boundary yields an integrable generalisation of the Kondo model with a gapped bulk.

Our approach is of course not completely unrelated to the traditional one. We go full circle and show our MPO for open boundary conditions can be written as a product of transfer matrices of the eight-vertex model, as also pointed out for the strong zero mode in the XXZ/six-vertex case \cite{Vernier2024}. Moreover, we show that the MPO defines an integrable model in its own right by explicitly constructing the $R$-matrix via the Yang-Baxter equation. 

In section 2, we construct the MPO and ensuing conserved charges for the XYZ chain for periodic and for free boundary conditions, In section 3,  we extend the calculation to include arbitrary boundary magnetic fields as well as impurities. There relations with traditional integrability approaches are discussed in section 4, while section 5 contains our conclusions.

\section{XYZ conserved charges from an MPO}

A simple MPO expression for conserved charges in the XYZ chain turns out to be rather straightforward to find, given the strong zero mode. We thus start our analysis by reviewing the result of \cite{Fendley2015} for the strong zero mode in the spin-$\tfrac12$ XYZ chain.  We rewrite it in terms of an elegant matrix product operator, and find a natural generalisation to a generating function for conserved charges.

% We show how to modify the boundary conditions in order to make this SZM an ESZM, i.e\ have it commute exactly with the Hamiltonian. 

\subsection{The strong zero mode as an MPO}

The original expression of the SZM for the spin-$\tfrac12$ XYZ chain was found by brute force \cite{Fendley2015}. One starts with the obvious fact that for $J_z\to\infty$ with $J_x$ and $J_y$ fixed, the operators $\sigma^z_j$ all commute with the Hamiltonian. The non-obvious fact is that one can start at one of the edges and iterate using the full Hamiltonian to find a sequence of operators $\Psi_r$ such that 
\begin{align}
\Psi(R)\equiv\sum_{r=1}^R \Psi_r,\qquad \Big[H_{\rm free}\,,\ \Psi(R)\Big] \sim \mathcal{O}(J_z^{-2R}) 
\label{HPsiXYZ}
\end{align}
for $R\le \lfloor L/2\rfloor$. Somewhat miraculously, the explicit expression for $\Psi_j$ is tractable. Defining the coupling ratios $K_x = \tfrac{J_z}{J_x}$, $K_y = \tfrac{J_z}{J_y}$, it is written in terms of the operator
\begin{align}
\psi(j,j')=K_y^{j'-j}\big(1-K_x^2\big)\sigma^x_{j}\sigma^x_{j'}+
K_x^{j'-j}\big(1-K_y^2\big)\sigma^y_{j}\sigma^y_{j'}\ ,
\label{psidef}
\end{align}
so that
\begin{align}
\Psi_r\equiv \sum_{0<j_1<\dots<j_{2r}<r}\big( K^{}_x K^{}_y\big)^{-(k-1)}\;\sigma^z_r\,\prod_{k=1}^r\,\psi\big(j_{2k-1},j_{2k}\big) \ .
\label{Psihalf}
\end{align}
%The localization near the ``close'' edge with $j$ small stems from the form of the expansion; a term with an operator $\sigma^a_{j}$ is multiplied by at least a power $1/J_z^{j-1}$.  
The operator $\Psi\big(\big\lfloor \tfrac{L}2\big\rfloor\big)$ involves all the spins, including that at the ``far'' edge with $j=L$. Truncating the sum in \eqref{HPsiXYZ} gives an operator commuting with the Hamiltonian up to terms coming from the far edge, which must be suppressed by at least a power of $J_z^{1-j}$. Thus as long as $K_x,K_y>1$, this correction becomes exponentially small far from the close edge.  If a coupling other than $J_z$ is largest, one simply swaps the couplings accordingly. 

%There does not appear to be any simple modification to $\Psi\big(\big\lfloor \tfrac{L}2\big\rfloor\big)$ that makes it commute exactly with the Hamiltonian. However, if we include a boundary magnetic field at the {\em far} edge, such a modification does exist \cite{Fendley2015}. Equivalently, we can include an extra spin at $j=L+1$, but do not include any terms in the Hamiltonian that can flip it, i.e.\ consider fixed boundary conditions for this spin. We write our expressions in the latter form, but one can obtain the former simply by setting $\sigma^z_{L+1}=1$. Including the field  allows us to convert the SZM into an ESZM, resulting in a conserved charge for the modified Hamiltonian. The Hamiltonian and the corresponding ESZM are given by
%\begin{align}
%H(h_L)=H^{\rm XYZ}_{\rm free} + b_L\,\sigma^z_L\sigma^z_{L+1}\ ,\quad 
%\Psi= \Psi\big(\big\lfloor \tfrac{L}2\big\rfloor\big) + \frac{J_z}{b_L}\big(K_x K_y\big)^{-2L}\Psi_{L+1}\ .
%\end{align}
%Proving that $\big[H(b_L),\,\Psi\big] = 0$ is a simple extension of the proof that $\Psi\big(\big\lfloor \tfrac{L}2\big\rfloor\big)$ is an SZM. The expression for $\Psi$ makes it clear that one cannot remove the magnetic field $b_L$ without destroying the ESZM. Worth noting is that the ESZM does not result in the pairing between energy levels like the SZM does; whether one interprets the modification as a boundary magnetic field or as a fixed boundary condition, it breaks the $\mathbb{Z}_2$ spin-flip symmetry.

While the explicit expression for the  brute-force method used for spin $\tfrac12$ seems unwieldy, it has a number of simple properties that make it ripe for rewriting as an MPO.  In any term in the sum \eqref{Psihalf}, $\sigma^z_j$ is always the rightmost operator. Crucially, the form of $\psi$ in \eqref{psidef} requires that the $\sigma^x_j$ always come in pairs with no operators in between, and likewise the $\sigma^y_j$. The MPO we need has four channels, i.e.\ is written in terms of 4$\times$4 matrices $A^a$. Since the strong zero mode is comprised entirely of Pauli matrices, we need four such matrices, letting $\sigma^0_j=1$. We thus consider an MPO of the form 
\begin{gather}
 \mathcal{M}_{k,k'} \equiv \sum_{\{a_j\},\{k_j\}} A^{a_1}_{kk_1}A^{a_2}_{k_1k_2} A^{a_3}_{k_2k_3}\dots A^{a_L}_{k_{L-1}k'}\,
\sigma^{a_1}_1 \sigma^{a_2}_2 \dots  \sigma^{a_L}_L\ .
\label{MPOdef}
\end{gather}
where we label both the matrix label $a$ and their indices by $0,x,y,z$. All sums in this paper not otherwise labelled are over $\{0,x,y,z\}$. It is then easy to check that the strong zero mode can be written as 
\begin{gather}
\label{PsiMPO}
\nonumber
\Psi\big(\big\lfloor \tfrac{L}2\big\rfloor\big)= \big(K_xK_y\big)^{-(L-1)}\mathcal{M}_{0,z},\\[0.2cm]
\hbox{with }\ A^0_{00}=A^z_{0z}=1,\qquad A^0_{bb}=\frac{J_bJ_z}{J_xJ_y},\qquad A^b_{0b}A^b_{b0}=A^0_{bb}- A^0_{cc}A^0_{dd}
\label{PsiA}
\end{gather}
where $b,c,d\in\{x,y,z\}$ such that $b\ne c\ne d\ne b$, a convention we maintain throughout this section.
All other matrix elements are zero, including $A^z_{z0}$. The latter vanishing is why the rightmost operator in each term of $ \mathcal{M}_{k,k'} $ must always be $\sigma^z_r$. Note as well that each side of the final relation in \eqref{PsiA} for $b=z$ indeed vanishes as it must. The freedom apparent for $b=x,y$ here arises because the $\sigma^x$ and $\sigma^y$ operators each always appear in pairs.

\subsection{Conserved charges for periodic and free boundary conditions}

The elegance of this MPO form strongly suggests that the strong zero mode is not a fluke, but that other conserved quantities (exact or almost) can be written in this fashion. The trick to finding this family is to demand that the MPO satisfy a certain {\em local} commutation relation, essentially that for a conserved current. Labelling the portion of the MPO acting on two consecutive sites as
\begin{align}\label{Mkkdef}
M^{(j,j+1)}_{kk'}\equiv \sum_{a,a',l} A^a_{kl}A^{a'}_{lk'}\,\sigma^a_j \sigma^{a'}_{j+1}\ ,
\end{align}
the conserved-current relation to be solved is  
\begin{align}
\left[h_{j,j+1},\,M^{(j,j+1)}_{kk'}\right] = \sum_{r,r',m} \left(E^{r}_{km} A^{r'}_{mk'} - A^{r}_{km} E^{r'}_{mk'}\right)\sigma^r_j \sigma^{r'}_{j+1}
\label{frank}
\end{align}
for any ``error'' terms $E^{a}_{kk'}$. This relation gives a set of bilinear relations constraining the $A$ and $E$ coefficients. It can be pictured as the following tensor diagram:
\begin{equation}
    \begin{tikzpicture}[baseline = 0cm]
    \path [draw = black] (-1,0) -- (1,0);
    \path [draw = black] (-0.5,-0.5) -- (-0.5,1.1);
    \path [draw = black] (0.5,-0.5) -- (0.5,1.1);
    \draw [draw = blue] (-0.5,0.3) rectangle (0.5,0.8);
    \node at (0,0.55) {$H$};
\end{tikzpicture}
\quad - \quad
\begin{tikzpicture}[baseline = 0cm]
    \path [draw = black] (-1,0) -- (1,0);
    \path [draw = black] (-0.5,-1.1) -- (-0.5,0.5);
    \path [draw = black] (0.5,-1.1) -- (0.5,0.5);
    \draw [draw =blue] (-0.5,-0.8) rectangle (0.5,-0.3);
    \node at (0,-0.55) {$H$};
\end{tikzpicture}
\quad = \quad
\begin{tikzpicture}[baseline = 0cm]
    \path [draw = black] (-1,0) -- (1,0);
    \path [draw = black] (-0.5,-0.5) -- (-0.5,0.5);
    \path [draw = black] (0.5,-0.5) -- (0.5,0.5);
    \draw [draw=green,fill = white] (-0.5,0) circle (0.25);
    \node at (-0.52,0) {$E$};
\end{tikzpicture}
\quad - \quad
\begin{tikzpicture}[baseline = 0cm]
    \path [draw = black] (-1,0) -- (1,0);
    \path [draw = black] (-0.5,-0.5) -- (-0.5,0.5);
    \path [draw = black] (0.5,-0.5) -- (0.5,0.5);
    \draw [draw=green,fill = white] (0.5,0) circle (0.25);
    \node at (0.48,0) {$E$};
\end{tikzpicture}
\end{equation}
where the vertices with four lines correspond to the matrix elements $A$, and the lines carry the labels.

An MPO satisfying \eqref{frank} for any choice of $k,k'$ immediately yields a conserved charge for the chain with periodic boundary conditions, as the error terms cancel pairwise:
\begin{align}
\Big[H_{\rm per},\, \hbox{tr}\, \mathcal{M}\Big] = 0\ ,\qquad\hbox{ for }\  
H_{\rm per}=\sum_{j=1}^L h_{j,j+1}\ ,\qquad  \hbox{tr}\, \mathcal{M}\equiv \sum_{k\in 0,x,y,z} \mathcal{M}_{k,k}\ ,
\label{HMcomm}
\end{align}
where the trace is over the ``internal" indices, not the Pauli matrices.
Moreover, when the error terms also satisfy $E^{a}_{0a'}=0=E^{a}_{a'0}$ for all $a$ and $a'$,
$[H_{\rm free},\mathcal{M}_{0,0}]=0$ as well. 

Matrices satisfying (\ref{frank}) and the ensuing conserved quantities for the XYZ spin chain can be found by generalising the strong zero mode in a very transparent fashion.  We impose the condition that for a matrix element $A^a_{bc}$ to be non-vanishing, it must have one index zero and the two others identical, so that the non-vanishing matrix elements are
\begin{equation}
A^0_{00},\quad A^b_{b0},\quad A^b_{0b},\quad A^0_{bb}\ .
\label{MPOform}
\end{equation}
We have thus generalised the form of \eqref{PsiA} to allow for $A^z_{0z}\ne 0$ as well. With this condition and normalizing $A^0_{00}=1$ , we find a one-parameter family of solutions of (\ref{frank}) given by
\begin{align}
\boxed{\quad 
A^0_{00}=1, \qquad A^0_{bb} = v J_b\ ,\qquad
A^b_{0b} A^b_{b0} =  v J_b - v^2 J_c J_d=A^0_{bb}- A^0_{cc}A^0_{dd}\ ,\quad}
\label{Acons}
\end{align}
where $b,c,d\in\{x,y,z\}$ such that $\epsilon_{bcd}\ne 0$, and $v$ is an arbitrary parameter.  It is simple to check that this MPO satisfies \eqref{frank}, with error terms given by
\begin{align} 
E^b_{cd} =  i\epsilon_{bcd}\, v^{-1}A^c_{c0} A^d_{0d}
\label{Edef}
\end{align}
with all others vanishing. 
In Appendix \ref{app:MPO} we prove that (\ref{Acons},\,\ref{Edef}) give the most general solution of \eqref{frank} with the constraints from \eqref{MPOform}.
One recovers the strong zero mode by setting $A^z_{0z}=0$, so that the last equation in \eqref{Acons} requires $v_{\rm SZM}=J_z/(J_x J_y)$ in this special case.

The MPO built from \eqref{Acons} thus gives a one-parameter family of operators commuting with the XYZ Hamiltonian with either periodic or free boundary conditions. From \eqref{HMcomm}, $\hbox{tr}\,\mathcal{M}$ commutes with $H_{\rm per}$, and since all error terms with a $0$ index vanish,  $\mathcal{M}_{0,0}$ commutes with $H_{\rm free}$. Thus allowing for non-vanishing $A^z_{0z}$ yields an exact symmetry, without the exponentially vanishing term for the strong zero mode as in  \eqref{HPsiXYZ}. The MPO here, however, is not localised at one of the edges. 

Since the MPOs commute with the corresponding Hamiltonian for any value of the parameter $v$, we can expand out the MPO in a power series. The coefficients are necessarily conserved quantities for the XYZ chain with either free or periodic boundary conditions. These MPOs are thus generating functions of conserved quantities whose number grows with $L$. We do need to check that these conserved quantities are non-trivial and so genuinely constrain the dynamics. 
Expanding the periodic MPO to order $v^3$ yields 
\begin{align}
\hbox{tr}\,\mathcal{M} = 1 + vH_{\rm per}+ v^2\Big( \tfrac12 H_{\rm per}^2 - L\big(J^2_x +J^2_y +J_z^2\big)\Big)
+ v^3\Big(H^{(3)} + \tfrac{1}{6} H_{\rm per}^3 - \alpha H_{\rm per}\Big)\ .
\label{MPOexp}
\end{align}
where $\alpha$ is an unimportant constant. Nothing new is generated at order $v^2$, as the MPO simply gives the square of the Hamiltonian plus a constant. Lest one worry that the MPO simply includes higher powers of the Hamiltonian, it is easy to check that the MPO is much simpler: taking powers of the Hamiltonian generates all sorts of terms forbidden by the very restrictive form of the MPO coming from \eqref{MPOform}. One also finds that if one defines $H^{(3)}$ via \eqref{MPOexp}, not only is it non-trivial, it is local. It thus can be thought of as a ``higher" commuting Hamiltonian. 

This conserved charge is precisely the one found using the traditional approach to integrability of spin chains. 
In this sophisticated but indirect approach, conserved quantities are found from an associated two-dimensional classical integrable model whose Boltzmann weights satisfy the Yang-Baxter equation. The spectral parameter $u$ is defined so that the resulting transfer matrices $T(u)$ obey $[T(u),\,T(u')]=0$. Expanding $T(u)$ in powers of $u$ yields then a hierarchy of conserved quantities commuting both with each other and the transfer matrix itself, just as we did with the MPO.  If a limit can be found such that one of these quantities is the Hamiltonian of the spin chain, then its integrability and the conserved quantities follow. For the XYZ chain, an associated classical model is the eight-vertex model, and the Boltzmann weights are given in terms of elliptic functions. Obviously, our approach is both elementary and more direct; nary an elliptic function is seen. It does not require introducing the full apparatus of integrability. Remarkably, despite considerable attention being given to the XYZ chain over the decades, such a simple expression for non-trivial conserved quantities does not seem to have been found before, save for the Heisenberg case \cite{Katsura2015}. 

Finding a one-parameter solution of \eqref{frank} gives an elementary derivation of the integrability of the XYZ chain. Worth noting is that our MPO cannot generate the analogous conserved charge $H^{(2)}$ coming from the transfer matrix: the latter is odd under spatial parity, while the MPO by construction is even. The two approaches, however, are closely related. We show in section \ref{sec:transfer} that the MPO for the open XYZ chain can be written in terms of a {\em product} of transfer matrices of the eight-vertex model. We thus expect that the MPO gives all even-parity higher conserved charges coming from the transfer matrix.

\section{Boundaries and impurities}

\subsection{Boundary magnetic fields}

The construction of the MPO conserved charges extends easily to having magnetic fields ${\mathfrak{h}}_b,\widetilde{\mathfrak{h}}_b$ acting on the boundary spins of an open chain. The resulting Hamiltonian generalises \eqref{HXYZ} to
\begin{equation}\label{Hopendef}
H_{\rm open}=H_{\rm free}+ \sum_{b\in\{x,y,z\}} \Big(\,\mathfrak{h}_b\, \sigma^b_1 + \widetilde{\mathfrak{h}}_b\, \sigma^b_L\,
\Big)\ .
\end{equation}
A natural generalisation of the MPO \eqref{MPOdef} to the boundary-field case is 
\begin{align}\label{Mopendef}
\mathcal{M}_{\rm open}=\sum_{\{a_j\},\{k_j\}} \mathcal{A}_{k_0}A^{a_1}_{k_0k_1}A^{a_2}_{k_1k_2} A^{a_3}_{k_2k_3}\dots A^{a_L}_{k_{L-1}k_L}\widetilde{\mathcal{A}}_{k_L}\,
\sigma^{a_1}_1 \sigma^{a_2}_2 \dots  \sigma^{a_L}_L\ ,
\end{align}
where the four-dimensional vectors $\mathcal{A}_k$ and $\widetilde{\mathcal{A}}_{k}$ are yet to be determined. 

Away from the boundaries, we require \eqref{frank} as before, so that the $A^a_{kk'}$ and the error terms remain as \eqref{Acons} and \eqref{Edef} respectively. The error terms in 
$[H_{\rm open},\,\mathcal{M}_{\rm open}]$ then still cancel telescopically in the bulk, leaving only contributions from the edges. We require that the error term for each edge cancels with the corresponding contribution of the field term to $[H_{\rm open},\,\mathcal{M}_{\rm open}]$, e.g.
\begin{align}\label{HA}
\Bigg[\sum_{b\in\{x,y,z\}} \mathfrak{h}_b \sigma^b_1\, ,\ \sum_{a_1,k_0}\mathcal{A}_{k_0}A^{a_1}_{k_0k_1}  \sigma^{a_1}_1 \Bigg]=-\sum_{r,k_0}\mathcal{A}_{k_0}E^{r}_{k_0k_1} \sigma^{r}_1
\end{align}
for the left edge. For each value of $k_1$ we have a matrix equation linear in the $\mathcal{A}_{k}$, as all else is known. Using \eqref{MPOform} and \eqref{Edef} yields
%\begin{equation}
%\begin{aligned}
\begin{align}
\label{hBA}
\mathfrak{h}_b\, \mathcal{A}_c A^c_{c0}=\mathfrak{h}_c \mathcal{A}_b A^b_{b0}\,, \qquad\quad
\mathfrak{h}_bA^c_{0c} \mathcal{A}_0=\frac{\mathcal{A}_b }{v}A^b_{b0}A^c_{0c}\,,
\end{align}
%\qquad k,\ell=x,y,z \quad k\neq\ell .
%\end{aligned}
%\end{equation}
where as before we take $b,c\in\{x,y,z\}$ with $b\ne c$. 
Non-vanishing boundary fields require $\mathcal{A}_0\neq 0$; for all $A^k_{k0}$ non-vanishing as well, the latter of \eqref{hBA} then implies the former. Then \eqref{HA} is satisfied for
\begin{align}
\mathcal{A}_b=v\frac{\mathfrak{h}_b}{A^b_{b0}}\mathcal{A}_0\,.
\end{align}
The equation for the right boundary follows by using $\widetilde{\mathfrak{h}}_b$ instead, and swapping the auxiliary (lower) indices in the $A$ and $E$ in equation \eqref{hBA}. The solution is 
\begin{align}
\widetilde{\mathcal{A}}_k=v\frac{\widetilde{\mathfrak{h}}_b}{A^b_{0b}}\widetilde{\mathcal{A}}_0\,.
\end{align}

The MPO $\mathcal{M}_{\rm open}$ therefore commutes with $H_{\rm open}$ for any choice of boundary fields $\mathfrak{h}_b,\widetilde{\mathfrak{h}}_b$. As with the periodic and free cases, it does so for any choice of the parameter $v$. Higher conserved local charges are thus generated by expanding in the free parameter $v$. For example, the Hamiltonian \eqref{Hopendef} can be simply obtained as the first derivative of the MPO at $v=0$:
\begin{align}
\label{Hderiv}
H_{\rm open}=\left.\frac{\rm d}{{\rm d}v}\right|_{v=0}\mathcal{M}_{\rm open}\,.
\end{align}
This elementary calculation therefore shows that the XYZ chain remains integrable for arbitrary boundary magnetic fields. 

\subsection{Impurities and Kondo with a gap}

The Yang-Baxter approach gives a method for deriving fine-tuned inhomogeneous couplings that preserve the integrability. Changing a coupling at a single location amounts to including an impurity. This analysis is rather unwieldy; in fact we are unaware of a closed-form expression for the impurity Hamiltonian in the full XYZ case. We give here a straightforward albeit slightly tedious derivation of the impurity Hamiltonian and the MPO generating the conserved charges. We show that if the impurity is placed at the edge, it gives a gapped lattice analog of the Kondo model. As with the uniform case above, we relegate some details to the appendix. 

We first consider an impurity at site $p$ somewhere in the bulk of the system, so that the interactions amongst the spins at $p-1,p,p+1$ are modified. We thus replace the interactions $h_{p-1,p}+h_{p,p+1}$ with $h_{p-1,p,p+1}$, as of yet unknown.
The simplest possibility for a modified MPO is to replace the entry $A^{a_p}_{k_pk_{p+1}}$ in \eqref{MPOdef} with a different set of matrices $D^{a_p}_{k_pk_{p+1}}$. For values of $j\ne p-1,p$, the usual relations \eqref{frank} apply. Those with $j=p-1,\,p$ are replaced with a single set of relations 
\begin{align}\label{Sascha_Eq}
\bigg[h^{}_{1,2,3},\,A^{a_{1}}_{k_{0},k_{1}} D^{a_{2}}_{k_{1},k_2}A^{a_3}_{k_2k_3}\sigma^{a_1}_1\sigma^{a_2}_2\sigma^{a_3}_3 \bigg]=\left(E^{r_1}_{k_0,k_1} D^{r_2}_{k_1,k_2}A^{r_3}_{k_2,k_3}-A^{r_1}_{k_0,k_1} D^{r_2}_{k_1,k_2}E^{r_3}_{k_2,k_3} \right)\sigma^{r_1}_1\sigma^{r_2}_2\sigma^{r_3}_3\,,
\end{align}
where for readability we took $p$\,=\,2 and omit writing the sums over all repeated indices.  These equations must be satisfied for all values of $k_0$ and $k_3$. 
Of course, taking $D=A$ and $h_{1,2,3}=h_{1,2}+h_{2,3}$ gives a solution.

While \eqref{Sascha_Eq} is a natural generalisation of the boundary relation \eqref{HA} to an impurity, a key distinction is that we do not {\em a priori} know a closed form for the impurity Hamiltonian $h_{p-1,p,p+1}$ that yields a solution; it is not arbitrary as with the boundary field. It is tempting to assume that the non-vanishing matrix elements of $D$ take the same form  \eqref{MPOform} for the uniform case.  We find that this does not quite work; we need to also allow for a non-vanishing $D^{b}_{cd}$, where we maintain the convention that $b,c,d\in\{x,y,z\}$ with $\epsilon_{bcd}\ne 0$. The non-vanishing matrix elements are therefore
\begin{align}
D^0_{00},\quad D^b_{b0},\quad D^b_{0b},\quad D^0_{bb},\quad D^b_{cd}\ .
%D^a_{a'0} \propto D^a_{0a'}\propto D^0_{aa'}\propto \delta_{aa'}\,.
\label{Dform}
\end{align} 
In addition, we look for an impurity Hamiltonian given by the first derivative of the MPO, just as in the uniform case. We thus impose
\[
h_{1,2,3}=\left.\frac{{\rm d}}{{\rm d}v}\right|_{v=0} \sum_{a_1,a_2,a_3} A^{a_1}_{0a_1}D^{a_2}_{a_1a_3}A^{a_3}_{a_3 0}\,\,\sigma^{a_1}_1\sigma^{a_2}_2\sigma^{a_3}_3\,
\equiv\sum_{a_1,a_2,a_3}h^{a_1 a_2 a_3}\,\, \sigma^{a_1}_1\sigma^{a_2}_2\sigma^{a_3}_3\ .
\]
With this assumption, for all $a_1,a_2,a_3$ the impurity Hamiltonian must obey
\begin{align}
\label{fsndjsndj}
h^{a_1 a_2 a_3}=\left. \frac{{\rm d}}{{\rm d}v}\right|_{v=0}  A^{a_1}_{0a_1}D^{a_2}_{a_1a_3}A^{a_3}_{a_3 0}\,.
\end{align}
If any upper index in \eqref{fsndjsndj} is set to be zero, then the other two must be equal to yield a non-vanishing right-hand side. If none are zero, our assumptions on $D$ imply that all need to be different. The non-vanishing components of the defect Hamiltonian are therefore analogous to those in $D$:
\begin{align}
h^{bb0}\,,\quad h^{b0b}\,,\quad h^{0bb}\,, \quad h^{bcd}\ , %\qquad k,\ell,m=x,y,z \qquad k\neq\ell,\quad k\neq m,\quad \ell\neq m
\label{hform}
\end{align}
where we can ignore $h^{000}$ because it multiplies the identity operator.

Even with these simplifying assumptions, the commutator in \eqref{Sascha_Eq} still involves many terms. Finding the solution is straightforward but a little tedious, so we include the details in Appendix \ref{app:impurity}. The impurity Hamiltonian has one free parameter $\alpha$, as follows from the traditional approach. The three-spin interaction is
\begin{align}
h^{c b d}&=\alpha\sqrt{\frac{J_c J_d}{J_b}-\alpha^2}\,,\qquad h^{d bc }=-s_bh^{c b d} \qquad\hbox{for }\ \epsilon_{bcd}=-1\ ,
\label{ham1}
\end{align}
while the $s_b$ are signs that obey:
\begin{align}
s_b=\pm 1 \,,\qquad \qquad s_xs_ys_z=1 \,.
\end{align}
The other contributions to the Hamiltonian are 
\begin{align}
h^{b0b} = \alpha^2 \, s_{b}\,,\qquad  h^{0bb}=\frac{h^{c d b}h^{d c b}}{h^{b0b}}\,,\qquad h^{bb0}=s_b\, h^{0bb}\ .
\label{ham2}
\end{align}
The MPO commuting with this Hamiltonian then is built using
\begin{equation}
\begin{gathered}
\label{Ddef}
D^0_{00}= 1-\alpha^2 v\,,  \qquad\quad D^0_{bb}=\frac{v}{A^0_{bb}}\Big( h^{b0b}+v\, \big(h^{0bb}h^{bb0}-\alpha^2h^{b0b}\big)\Big)\ ,\cr
D^b_{b0}=h^{bb0}\frac{A^b_{b0}}{A^0_{bb}}\, v \, ,\qquad\quad D^b_{0b}=h^{0bb}\frac{A^b_{0b}}{A^0_{bb}}\,v\,,\qquad\quad D^{b}_{c d}= h^{c b d}\frac{A^{c}_{c0}A^{d}_{0d}}{A^0_{cc}A^0_{dd}}\ .
\end{gathered}
\end{equation}
Taking the limit $\alpha\to 0$ yields the uniform Hamiltonian and MPO.

Our MPO approach thus shows how to include impurities into the XYZ chain while preserving the integrability, yielding the three-site Hamiltonian $h_{p-1,p,p+1}$ with the fine-tuned form (\ref{ham1},\ref{ham2}). Our derivation, following from \eqref{Sascha_Eq}, shows that impurities of this form can be placed at any points $p_l$, as long as $|p_l-p_{l'}|>1$. 

A very interesting system results from placing the impurity at the edge, i.e.\ taking $p$\,=\,1. This chain is a lattice version of the integrable Kondo problem, where a fluctuating spin is placed at the end of a system of fermions (see e.g.\ \cite{Saleur1998} and references therein). The conventional Kondo problem has a gapless bulk, but as the XYZ chain has a gap in general, the system thus provides a gapped analog. We first consider the case where we preserve both $\mathbb{Z}_2$ symmetries, and so do not include any boundary term in the MPO. Simply placing the $D$ term at the end, we look for a solution of 
\begin{align}
\sum_{k_1,a_1,a_2}\bigg[K_{1,2},\;D^{a_{1}}_{{0}k_1}A^{a_2}_{k_1k_2}\sigma^{a_1}_1\sigma^{a_2}_2 \bigg]=- \sum_{k_1,r_1,r_2}D^{r_1}_{0k_1}E^{r_2}_{k_1k_2}\sigma^{r_1}_1\sigma^{r_2}_2
\end{align}
for all $k_2$.
As before, to match the bulk, the error terms must remain the same in the presence of a boundary and/or an impurity.
Assuming $D$ remains as in \eqref{Ddef}, it follows easily that the boundary Hamiltonian is
\begin{align}
K_{1,2} = \sum_{b\in\{x,y,z\}} h^{0bb}\sigma^b_1\sigma^b_2\ .
\end{align}
As $\alpha\to 0$, $h^{0bb}$ indeed reduces to $J_b$, but in general it is rather non-obvious.

To allow for a boundary field breaking the symmetry, we need to combine the ideas behind \eqref{HA} and \eqref{Sascha_Eq} and consider an MPO of the form $\mathcal{A}DA$ at the left edge. The matrix elements must satisfy
\begin{align}\label{Kondo}
\sum_{k_0,k_1,a_1,a_2}\bigg[\mathcal{K}_{1,2},\;\mathcal{A}_{k_{0}} D^{a_{1}}_{k_{0}k_1}A^{a_2}_{k_1k_2}\sigma^{a_1}_1\sigma^{a_2}_2 \bigg]=-
\sum_{k_0,k_1,r_1,r_2}\mathcal{A}_{k_0} D^{r_1}_{k_0k_1}E^{r_2}_{k_1k_2}
%-E^{r_1}_{k_0,k_1} D^{r_2}_{k_1,k_2}A^{r_3}_{k_2,k_3} \right)
\sigma^{r_1}_1\sigma^{r_2}_2\,.
\end{align}
where the Hamiltonian $\mathcal{K}_{1,2}$ encompasses both the boundary fields and the neighbouring impurity. Assuming that the MPO constituents are the same as above, solving \eqref{Kondo} gives 
\begin{align}
\mathcal{K}_{1,2}= K_{1,2} +\sum_{b\in\{x,y,z\}} \frac{\mathfrak{h}_b}{J_b}\left( h^{bb0}\sigma^b_1+h^{b0b}\sigma^b_2 
+ h^{bcd}\sigma^{c}_1\sigma^{d}_2 + h^{bdc}\sigma^{d}_1\sigma^{c}_2 \right)\ .
\end{align}

\section{Relation to traditional approaches }

In this section we make contact with the traditional integrability approach in several ways. We show how to write the conserved-charge MPOs in terms of a product of two transfer matrices of the eight-vertex model. The fact that the MPO yields only the parity-invariant conserved charges thus becomes less mysterious; those odd under parity cancel in the product. Given that that the MPO can be written in terms of transfer matrices of an integrable model, it is natural to expect that it defines an integrable model in its own right. We indeed demonstrate its integrability directly by finding the corresponding $R$ matrix.

%%%%%%%%%%%%%%%%%%%%%%%%%%%%%%%%%%%%%%%%%%%%%%%%%%%%%%%%%%%%%%%%%%%%%%%%
\subsection{The MPO as a product of transfer matrices}
\label{sec:transfer}
%%%%%%%%%%%%%%%%%%%%%%%%%%%%%%%%%%%%%%%%%%%%%%%%%%%%%%%%%%%%%%%%%%%%%%%%

The basic building blocks of these transfer matrices are defined by the operators
\begin{align}
g_{j,j+1}  = \sum_{a\in\{0,x,y,z\}} w_a \sigma^a_{j}\sigma^a_{j+1}\ .
\label{gdef}
\end{align}
The $w_a$ are as of yet undetermined; they will turn out to be related to the eight-vertex model Boltzmann weights. We define the product of $L$ of these operators acting on $L+1$ sites as
\begin{align}
T_{L+1}=g_{L,L+1}g_{L-1,L}\dots g_{1,2}\ .
\label{Tg}
\end{align}
and then form the product
\begin{align}
\overline{\cal M}_k(L+1) \equiv T_{L+1}\, \sigma^k_1\, \big(T_{L+1})\big)^{\rm T}\ .
\label{MTT}
\end{align}
where ${\rm T}$ here means transpose; notice that $g^{\rm T}_{j,j+1}=g_{j,j+1}$. 

This product operator $\overline{\cal M}_a$ is related to the MPO by decomposing it into a tensor product of its action on first $L$ sites and the $(L+1)$st via
\begin{align}
\overline{\cal M}_k(L+1) = \sum_{k'} 
\mathcal{M}_{k,k'}\otimes\sigma^{k'}_{L+1}
\label{MMM}
\end{align}
where $\mathcal{M}_{k,k'}$ by definition acts only on the first $L$ sites. We prove that $\mathcal{M}_{k,k'}$ is precisely that defined by \eqref{MPOdef} and \eqref{Acons} once the coefficients are identified appropriately. The proof utilises two easily established identities. One is 
\begin{align}
\big(g_{j,j+1}\big)^2= \sum_a w_a^2\ +\ 2\sum_{b\in\{x,y,z\}} (w_0w_b-w_cw_d) \sigma^b_j\sigma^b_{j+1}\ .
\label{ident2}
\end{align}
where we maintain the convention that $b,c,d\in\{x,y,z\}$ with $\epsilon_{bcd}\ne 0$.  The other identity is
\begin{align}
g_{j,j+1}\, \sigma^b_{j}\, g_{j,j+1} = \big(w_0^2+w_b^2-w_c^2-w_d^2\big)\sigma^b_{j} +2\big(w_0w_b+w_cw_d\big) \sigma^b_{j+1} \ ,
\label{ident1}
\end{align}

We now prove \eqref{MMM} recursively in $L$.  First consider $L$\,=\,1, where \eqref{MPOdef} requires that
$\mathcal{M}_{k,k'}(1)= \sum_a A_{kk'}^a\sigma^a_1$, while by definition $\overline{\cal M}_k(2)=g_{1,2}\,\sigma^k_1\,g_{1,2}$. For \eqref{MMM} with $L$\,=\,1 to hold with $k$\,=\,0, the MPO coefficients therefore must be related to the $w_a$ by
\begin{align}
A_{00}^0 = \sum_a w_a^2\ ,\qquad
A_{0b}^b = 2(w_0w_b-w_cw_d)\ ,
\label{A0b}
\end{align}
using \eqref{ident2}.
For it to hold for the other values of $k$, using (\ref{ident1}) yields
\begin{align}
A_{b0}^b = w_0^2+w_b^2-w_c^2-w_d^2\ ,\qquad
A_{bb}^0 = 2(w_0w_b+w_cw_d)\ .
\label{Abb}
\end{align}
Requiring that \eqref{MMM} holds for $L=1$ thus fixes all the coefficients and lets us rewrite the identities (\ref{ident2},\ref{ident1}) in terms of the MPO coefficients.
To prove \eqref{MMM} for all $L$ by recursion, assume that it holds for $\overline{\cal M}_k(L+1)$. Using the definition \eqref{MTT} along with the rewritten identities, we have
\begin{align*}
\overline{\cal M}_k(L+1)&=g_{L+1} \left(\overline{\cal M}_k(L+1)\otimes 1 \right)g_{L+1}\cr 
&= %\big(\mathcal{M}_{a,0}\otimes 1\otimes 1\big) g_{L+1}^2 + 
\sum_{k'} \big(\mathcal{M}_{k,k'}\otimes 1\otimes 1\big) g_{L+1} \sigma^{k'}_{L+1} g_{L+1}\cr
&=(\mathcal{M}_{k,0}\otimes 1\otimes 1) \sum_{a}A_{0a}^{a} \sigma^{a}_{L+1}\sigma^{a}_{L+2}\ +\  \sum_{b\in\{x,y,z\}} (\mathcal{M}_{k,b}\otimes 1\otimes 1) \left( A_{b0}^b \sigma^b_{L+1} + A_{bb}^0 \sigma^b_{L+2}\right)\cr
&=\sum_{k'} (\mathcal{M}_{k,k'}^{(L+1)}\otimes 1) \sigma^{k'}_{L+2}
\end{align*}
where $\mathcal{M}_{k,k'}^{(L+1)}$ is the MPO on $L+1$ sites. This relation is indeed \eqref{MMM} for $L+1\to L+2$ sites.

We thus have shown that the product of transfer matrices $\overline{\mathcal{M}}_{k}(L+1)$ from \eqref{MTT} indeed can be expanded in terms of the MPO $\mathcal{M}_{k,k'}$ as in \eqref{MMM}, if we write the MPO coefficients in terms of the $w_a$ as in (\ref{A0b},\,\ref{Abb}). For this MPO to yield conserved charges, these coefficients must obey the conditions in \eqref{Acons}. The XYZ couplings therefore must be related to the $w_a$ as
\begin{align}
\frac{A_{bb}^0}{
A_{00}^0} = J_b v =  \frac{ 2(w_0w_b+w_cw_d)}{w_0^2+w_x^2+w_y^2+w_z^2}\ .
\label{Au}
\end{align}
A little algebra then shows that the expressions (\ref{A0b},\ref{Abb}) do indeed satisfy the last relation in \eqref{Acons}.

The product $T_{L+1}$ is a transfer matrix for the eight-vertex model on the square lattice with open boundary conditions, as follows from \cite{Baxter1982}, section 10.14. Its matrix elements can be pictured by a graph where each edge corresponds to a two-state system with the corresponding label $j=1,2,\dots L+1$, namely
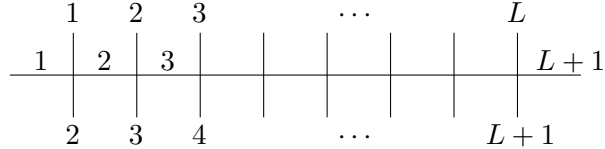
\begin{center}
\begin{picture}(220,55)
\put(9,28){1}
\put(0,26){\line(1,0){216}}
\put(21,0){2}
\put(24,10){\line(0,1){32}}
\put(21,46){1}
\put(33,28){2}
\put(48,10){\line(0,1){32}}
\put(45,46){2}
\put(45,0){3}
\put(72,10){\line(0,1){32}}
\put(57,28){3}
\put(69,46){3}
\put(69,0){4}
\put(96,10){\line(0,1){32}}
\put(120,10){\line(0,1){32}}
\put(144,10){\line(0,1){32}}
\put(168,10){\line(0,1){32}}
\put(192,10){\line(0,1){32}}
\put(124,0){$\cdots$}
\put(124,46){$\cdots$}
\put(180,0){$L+1$}
\put(188,46){$L$}
\put(199,28){$L+1$}
\end{picture}
\end{center}
The Pauli matrices $\sigma^a_j$ map the lower set of two-state systems to the upper set. The transfer matrix \eqref{Tg} then acts up and to the right, starting with $g_1$ and with the others in succession. Each matrix element of $T_{L+1}$ corresponds to a particular state on each edge. 

To make contact with the eight-vertex model, we label the state on each edge of this graph with an arrow such that $\sigma^z=1$ when it points up or to the right, and $\sigma^z=-1$ when it points down or to the left. Each of the $2^{L+1}\times 2^{L+1}$ elements of $T_{L+1}$ corresponds to a choice of arrows on the 2$L$\,+\,1 edges. Not all choices occur; since $g_j$ only flips spins in pairs, there are only eight possible choices at each vertex, corresponding to an even number of arrows in. Each such choice corresponds to a configuration of the classical eight-vertex model. The Boltzmann weights of the integrable eight-vertex model are the product of weights depending on the configuration at each vertex. These weights are conventionally labelled $a,b,c,d$ (not to be confused with our labelling of matrix indices!) so that
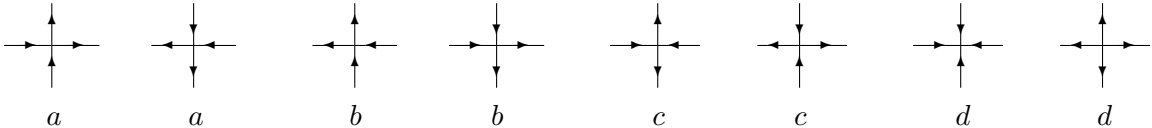
\[
\begin{picture}(50,50)
\put(0,30){\vector(1,0){12}}\put(6,30){\line(1,0){12}}
\put(18,14){\vector(0,1){12}}\put(18,26){\line(0,1){4}}
\put(16,0){$a$}
\put(18,30){\vector(1,0){12}}\put(24,30){\line(1,0){12}}
\put(18,30){\vector(0,1){12}}\put(18,42){\line(0,1){4}}
%\put(72,16){$\longrightarrow$}
%
\end{picture}
\ 
\begin{picture}(50,50)
\put(18,14){\line(0,1){4}}\put(18,30){\vector(0,-1){12}}
\put(2,30){\line(1,0){4}}\put(18,30){\vector(-1,0){12}}
\put(16,0){$a$}
\put(18,30){\line(0,1){4}}\put(18,46){\vector(0,-1){12}}
\put(18,30){\line(1,0){4}}\put(34,30){\vector(-1,0){12}}
\end{picture}
\quad
\begin{picture}(50,50)
\put(18,14){\vector(0,1){12}}\put(18,26){\line(0,1){4}}
\put(16,0){$b$}
\put(18,30){\vector(0,1){12}}\put(18,42){\line(0,1){4}}
\put(2,30){\line(1,0){4}}\put(18,30){\vector(-1,0){12}}
\put(18,30){\line(1,0){4}}\put(34,30){\vector(-1,0){12}}
%\put(72,16){$\longrightarrow$}
%
\end{picture}
\
\begin{picture}(50,50)
\put(18,14){\line(0,1){4}}\put(18,30){\vector(0,-1){12}}
\put(18,30){\line(0,1){4}}\put(18,46){\vector(0,-1){12}}
\put(16,0){$b$}
\put(0,30){\vector(1,0){12}}\put(6,30){\line(1,0){12}}
\put(18,30){\vector(1,0){12}}\put(24,30){\line(1,0){12}}
\end{picture}
\quad
\begin{picture}(50,50)
\put(18,14){\line(0,1){4}}\put(18,30){\vector(0,-1){12}}
\put(0,30){\vector(1,0){12}}\put(6,30){\line(1,0){12}}
\put(16,0){$c$}
\put(18,30){\vector(0,1){12}}\put(18,42){\line(0,1){4}}
\put(18,30){\line(1,0){4}}\put(34,30){\vector(-1,0){12}}
%\put(72,16){$\longrightarrow$}
%
\end{picture}
\ 
\begin{picture}(50,50)
\put(18,14){\vector(0,1){12}}\put(18,26){\line(0,1){4}}
\put(2,30){\line(1,0){4}}\put(18,30){\vector(-1,0){12}}
\put(16,0){$c$}
\put(18,30){\line(0,1){4}}\put(18,46){\vector(0,-1){12}}
\put(18,30){\vector(1,0){12}}\put(24,30){\line(1,0){12}}
\end{picture}
\quad 
\begin{picture}(50,50)
\put(0,30){\vector(1,0){12}}\put(6,30){\line(1,0){12}}
\put(18,14){\vector(0,1){12}}\put(18,26){\line(0,1){4}}
\put(16,0){$d$}
\put(18,30){\line(0,1){4}}\put(18,46){\vector(0,-1){12}}
\put(18,30){\line(1,0){4}}\put(34,30){\vector(-1,0){12}}
\end{picture}
\
\begin{picture}(50,50)
\put(18,14){\line(0,1){4}}\put(18,30){\vector(0,-1){12}}
\put(2,30){\line(1,0){4}}\put(18,30){\vector(-1,0){12}}
\put(16,0){$d$}
\put(18,30){\vector(1,0){12}}\put(24,30){\line(1,0){12}}
\put(18,30){\vector(0,1){12}}\put(18,42){\line(0,1){4}}
\end{picture}
\]
The invariance of the weights under reversing all arrows is known here as the ``zero-field" condition, and it results in the eight-vertex model having the same pair of $\mathbb{Z}_2$ symmetries as the XYZ Hamiltonian.

The transfer matrix for this eight-vertex model is simply $T_{L+1}$, once we appropriately identify the Boltzmann weights $a,b,c,d$ with those in the XYZ chain. To make this identification, we note that each $g_{j,j+1}$ maps the left and bottom states to the upper and right states. For example, 
\begin{center}
\begin{picture}(160,32)
\put(-45,12){$X_jX_{j+1}:$}
\put(20,16){\vector(1,0){12}}\put(26,16){\line(1,0){12}}
\put(38,0){\vector(0,1){12}}
\put(38,12){\line(0,1){4}}
\put(72,16){$\longrightarrow$}
\put(128,16){\line(0,1){4}}
\put(128,32){\vector(0,-1){12}}
\put(128,16){\line(1,0){4}}
\put(144,16){\vector(-1,0){12}}
\end{picture}
\end{center}
%\begin{align}
%\begin{picture}(200,32)
%\put(0,12){$d=$}
%\put(20,16){\vector(1,0){12}}\put(26,16){\line(1,0){12}}
%\put(38,0){\vector(0,1){12}}
%\put(38,12){\line(0,1){4}}
%\put(38,16){\line(0,1){4}}
%\put(38,32){\vector(0,-1){12}}
%\put(38,16){\line(1,0){4}}
%\put(54,16){\vector(-1,0){12}}
%\end{picture}
%\end{align}
The eight-vertex transfer matrix is then indeed $T_{L+1}$ from \eqref{Tg}, where
$g_j$ is defined by \eqref{gdef} with
\begin{align}
w_0 = \frac{a+c}{2}\ ,\quad w_z = \frac{a-c}{2}\ ,\quad
w_x = \frac{b+d}{2}\ ,\quad w_y = \frac{b-d}{2}\ .
\end{align}
The XYZ Hamiltonian can be obtained from the transfer matrix by noting that $T_{L+1}=1$ when $a$\,=\,$d$\,=1 and $b$\,=\,$c$\,=0.  Varying the couplings slightly away from this limit, one then finds 
\begin{align}
H_{\rm free}\ \propto\ T -1
\end{align}
as long as we identify
\begin{align}
\frac{J_b}{J_c} = \frac{w_0w_b+w_cw_d}{w_0w_c+w_bw_d}
\label{JJ}
\end{align}
just as in (\ref{Au}).  The MPO $\mathcal{M}_{k,k'}$ and hence the conserved charges are indeed related to a product of eight-vertex model transfer matrices in a very natural fashion.

%%%%%%%%%%%%%%%%%%%%%%%%%%%%%%%%%%%%
\subsection{The \texorpdfstring{$R$}{R} matrix}
\label{sec:Rmatrix}

Since the conserved-charge MPO can be expressed in terms of products of transfer matrices of an integrable model, it is natural to expect that it defines the transfer matrix of an integrable model in its own right. We find that indeed 
\begin{align}
\big[\hbox{tr}\,\mathcal{M}(v),\,\hbox{tr}\,\mathcal{M}(v')\big]\ =\ 0\,.
\label{MMcomm}
\end{align}

The proof follows from implementing the standard Yang-Baxter procedure. We find a 16$\times$16 matrix $R(v,v')$ that intertwines two MPOs with different spectral parameters $v$ and $v'$, namely 
\begin{align}
\sum_{l,l'} R^{k',l'}_{k,l}(v,v') \mathcal{M}_{l',m'}(v') \mathcal{M}_{l,m}(v) = \sum_{l,l'} \mathcal{M}_{k',l'}(v)  \mathcal{M}_{k,l}(v') R^{l',m'}_{l,m}(v,v')\ .
\label{RMM}
\end{align}
The commutation relation \eqref{MMcomm} follows immediately from \eqref{RMM} as long as the $R$ matrix is invertible. The $R$ matrix is found by requiring it solve the linear equations for $\mathcal{M}$ with $L$\,=\,1, that is, pictorially,
\begin{align} 
\sum_{l,l'}
\quad
\begin{tikzpicture}[baseline=(current  bounding  box.center),scale=1.]
\draw[rounded  corners=10pt] (0,-0.5) -- (0.5,-0.5) -- (1,0);
\node at (0.25,0.75) {$ k' $};
\node at (0.25,-0.25) {$ k$};
\node at (1.65,0.75) {$ l' $};
\node at (1.65,-0.25) {$ l$};
\node at (2.35,0.75) {$ m' $};
\node at (2.35,-0.25) {$ m$};
\draw[ rounded  corners=10pt] (1,0) -- (1.5,0.5) -- (2.5,0.5);
\draw[rounded  corners=10pt] (0,0.5) -- (0.5,0.5) -- (1,0);
\draw[ rounded  corners=10pt] (1,0) -- (1.5,-0.5) -- (2.5,-0.5);
\draw (2,-1) -- (2,1);
\end{tikzpicture}
\quad 
=
\quad
\sum_{l,l'}
\quad
\begin{tikzpicture}[baseline=(current  bounding  box.center),scale=1.]
\node at (0.55,0.75) {$ k' $};
\node at (0.55,-0.25) {$ k$};
\node at (1.45,0.75) {$ l' $};
\node at (1.45,-0.25) {$ l$};
\node at (2.6,0.75) {$ m' $};
\node at (2.6,-0.25) {$ m$};
\draw[rounded  corners=10pt] (0.5,-0.5) -- (1.5,-0.5) -- (2,0);
\draw[rounded  corners=10pt] (2,0) -- (2.5,0.5) -- (3.,0.5);
\draw[rounded  corners=10pt] (0.5,0.5) -- (1.5,0.5) -- (2,0);
\draw[rounded  corners=10pt] (2,0) -- (2.5,-0.5) -- (3.,-0.5);
\draw (1,-1) -- (1,1);
\end{tikzpicture}
\label{eq:YBE}
\end{align} 
Applying this relation repeatedly down the line gives \eqref{RMM}. 
Up to a global normalization, \eqref{eq:YBE} entirely fixes the coefficients $R_{k,l}^{k',l'}$. 
Because of the MPO form \eqref{Acons}, we find the coefficients are non-vanishing in the cases where all indices coincide, where they coincide two by two, and where they are all different. 

The calculation of $R$ is then straightforward, so we simply present the answer.  We choose a normalization such that 
\begin{align} 
R_{k,k}^{k,k}(v,v') = 1 
\end{align} 
for all $k \in \{0,x,y,z\}$. 
To fix the remaining entries, it is convenient to introduce the ``gauge independent'' matrix $S_{k,l}^{k',l'}(v,v')$ defined by :  
\begin{align}
R_{k,l}^{k',l'}(v,v') = A_{k0}^k(v')A_{k'0}^{k'}(v) 
A_{0l}^l(v) A_{0l'}^{l'}(v') 
S_{k,l}^{k',l'}(v,v') 
\end{align}
Keeping the convention that indices $b,c,d \in \{x,y,z\}$ with $\epsilon_{bcd}\neq 0$, we find
\begin{equation}
\begin{gathered}
S_{0,b}^{d,c}= i \frac{\mathcal{S}(v,v')}{J_d v} \epsilon_{bcd}\,,\quad 
S_{c,0}^{b,d} = i \frac{\mathcal{S}(v,v')}{J_d v'} \epsilon_{bcd}\,,\quad
S_{b,d}^{c,0} = i \frac{\mathcal{S}(v,v')}{J_d v} \epsilon_{bcd}\,,\quad
S_{d,c}^{0,b} = i \frac{\mathcal{S}(v,v')}{J_d v'} \epsilon_{bcd}
\cr
S_{b,b}^{0,0}=S_{0,0}^{b,b}= 
\frac{\big(J_d-v J_b J_c\big)\big(J_c-u J_d J_b\big)+\big(J_d-u J_b J_c\big)\big(J_c-v J_d J_b\big)}{J_a J_b J_c (v'-v)} \mathcal{S}(v,v')
\cr
 \big(v' S_{b,0}^{0,b} - v  S_{b,b}^{0,0} \big)\big(J_b-v' J_c J_d\big) = \big(v S_{0,b}^{b,0} - v'  S_{b,b}^{0,0} \big)\big(J_b-v J_c J_d\big) = 1 
\cr 
S_{b,c}^{c,b}  = \frac{S_{0,c}^{c,0} }{v'\big(J_b- v' J_c J_d\big)} + \frac{\mathcal{S}(v,v')}{v v'} \frac{J_d - v' J_b J_c}{J_b- v' J_c J_d}\,,\quad\ 
S_{b,b}^{c,c}=   \frac{J_d(v'+v)-2 v v' J_b J_c}{J_bJ_c J_d(v'-v)}\mathcal{S}(v,v') \,,
\label{Sdef}
\end{gathered}\end{equation}
where the scattering factor $\mathcal{S}(v,v')$ is given by
\[ 
\frac{1}
{\mathcal{S}(v,v')}=
\frac{v+v'}{v'-v}\left(
 \left(v v' \left(J_x^2+J_y^2+J_z^2\right)+1\right)-
   J_x J_y J_z \left(2\tfrac{v^2 v'^2}{v+v'}+\tfrac{1}{2} (v+v')
 \left(\tfrac{1}{J_x^2}+\tfrac{1}{J_y^2}+\tfrac{1}{J_z^2}\right)\right)\right) \,.
\]
For ease of presentation we have omitted the $v,v'$ arguments in the matrix elements of $S$ in \eqref{Sdef}.

Given the interpretation of the MPO as a product of eight-vertex transfer matrices, the $R$ matrix written above could also be reconstructed as a product of four $R$ matrices of the original eight-vertex model. As a consequence it satisfies the Yang-Baxter equation, as we checked explicitly. Parametrizing the eight-vertex weights in the usual elliptic fashion in terms of additive spectral parameters $u,u'$, the weights entering the definition of $R$ have for arguments $u'-u$, $u'+u$, $u-u'$, $-u-u'$. It is therefore unsurprising that $R$ cannot be brought to a difference form, even though the original eight-vertex $R$ matrices are.

%\newpage
\section{Conclusion}

Often advances in science have been “in the air”, so that it was only a matter of time before someone made the necessary leap. Baxter’s profound work on the eight-vertex model came from a very different place:  both results and methods introduced were singular. As a consequence, we are still coming to grips with some of their implications. These results remain extremely difficult to generalise to other non-critical lattice models. While many ``elliptic" integrable models without a U(1) symmetry are now known, the main reliable (albeit mysterious) technique for analysing their physics is via the corner-transfer matrix approach \cite{Baxter1982}. 

The results in this paper point to a potentially fruitful approach to this problem. Deriving the Bethe equations for our MPO and hence the XYZ chain might be more straightforward than going through the full eight-vertex model analysis. Intriguingly, there has been renewed interest in formulating a more elementary version of the Bethe ansatz for the XYZ model directly. In particular, significant progress in this direction has resulted from a chiral Bethe ansatz in which the elementary excitations correspond to kinks in a chiral helical basis \cite{Zhang:2022atz,Zhang2023}. We have found that our MPO splits some of the degeneracies of the spin chain, giving hope that XYZ and other non-U(1)-invariant chains may allow their integrability to be exploited in a new fashion. 

At minimum, our results make the integrability of the XYZ Hamiltonian less mysterious. Moreover, our results provide a direct connection between the strong zero mode and integrability, albeit not via the traditional techniques of integrability such as the Bethe ansatz. Indeed, our results are closely related to recent progress made on connecting the strong zero mode with integrability \cite{Vernier2024, essler2025strong, Gehrmann2025}.

Our results may also prove useful in finding and classifying integrable models. Recent works on systematically finding new integrable models are for example 
\cite{Corcoran2024,deLeeuw2024}.
A promising method for classifying integrable models combines two of the traditional approaches; see \cite{deLeeuw2020}
and references therein for an overview. It makes use of the Sutherland equation, which is obtained by differentiating the Yang–Baxter equation, along with the boost operator \cite{Tetelman1982, 10.1143/PTP.70.1508, Links2001}, which generates the tower of higher conserved charges from an integrable Hamiltonian density. In this approach, one first solves for a candidate Hamiltonian density by requiring the existence of a third non-trivial conserved charge generated by the boost. The candidate density is then fed into the Sutherland equation, which in turn one attempts to solve. In contrast to this, our equation \eqref{frank} is more general than the Sutherland equation and can be solved directly, without the need to find an {\em a priori} connection with some classical integrable model.

There are interesting complementary results on XYZ conserved charges. In the XXZ case, an elegant closed-form expression for the conserved charges in terms of the Temperley-Lieb generators has been found \cite{Nienhuis2021}. This expression can be generalised to the XYZ case \cite{Nozawa2020}, but it is not as simple as the MPO here. However, in unpublished work \cite{Fukai2025}, a simple MPO expression for the homogenous case is found that also includes the odd-parity conserved charges. The story begun by Baxter more than half a century ago is thus  far from complete.

%%%%%%%%%%%%%%%%%%%%%%%%%%%%%%%%%%%%
\section*{Acknowledgements}
PF and SG were supported by the EPSRC under grant EP/X030881/1. 
EV was supported by the ANR under grant ANR-24-CE40-7252. FV acknowledges funding from the  UKRI grant EP/Z003342/1 and from EOS (Grant No. 40007526), IBOF (Grant No. IBOF23/064), and BOF-GOA (Grant No. BOF23/GOA/021). We thank Fabian Essler, Akshat Pandey and Weronika Wiesiolek for stimulating discussions.

\begin{appendix}

\appendix
\section{Derivation of the MPO with uniform couplings}
\label{app:MPO}

Here we explicitly find the most general solution \eqref{Acons} of the relation \eqref{frank}, given the definition \eqref{MPOdef} with the form \eqref{MPOform}. We keep the convention that any index can take values $0,x,y,z$, except for $b,c,d\in\{x,y,z\}$, such that $\epsilon_{bcd}\ne 0$. We parametrise the matrix elements as
\begin{align}
A^0_{00}=1\ ,\qquad A^{0}_{bb}=  B_b\ , \qquad A^{b}_{0b}=  C_b\ ,\quad  A^{b}_{b0}=  \widetilde{C}_b\ ,
\label{ABC}
\end{align}
with all others vanishing. For convenience, we take $C_0=\widetilde{C}_0=B_0=0=J_0$.
%and sum over all repeated indices unless indicated otherwise. 

There are three types of terms in the commutator in \eqref{frank} those proportional to $I\otimes \sigma^b$, $\sigma^b\otimes I$ and $\sigma^b\otimes \sigma^c$. Those of the first type have coefficient
\[ \sum_{a,a',l,c} i\epsilon_{ca'b}J_c\delta_{ca}A^a_{kl}A^{a'}_{lk'} =  \sum_{a',c}  i\epsilon_{ca'b}J_c\,\delta_{ck}C_c\widetilde{C}_{a'}\delta_{k'a'} 
= i\epsilon_{kk'b} J_k C_k \widetilde{C}_{k'} \ .
\]
When the right-hand side is non-vanishing \eqref{frank} yields
\begin{align}
-B_c E^b_{ck'}=  i\epsilon_{ck'b} J_c \widetilde{C}_c C_{k'}\ ,
\label{constraint1}
\end{align}
so that $E^b_{bb}=E^b_{cc}=E^b_{b0}=E^b_{c0}=0.$ We have relabelled $k\to c\in\{x,y,z\}$ here to emphasise the fact that this equation provides useful information only for $k\ne 0$. 
Repeating this calculation gives the coefficient of the $\sigma^b\otimes I$ term as
\[  \sum_{a,a',l,c} i\epsilon_{cab}J_c\delta_{ca'}A^a_{kl}A^{a'}_{lk'} =   \sum_{a,c}i\epsilon_{cab}J_c\,C_a\delta_{ka}\widetilde{C}_{a'}\delta_{k'c} = i\epsilon_{k'kb} J_r C_k \widetilde{C}_{k'} \ .\]
so that
\begin{align}
 B_{d} E^b_{kd} = i\epsilon_{dkb} J_{d}  \widetilde{C}_k C_{d}
\label{constraint2}
\end{align}
along with $E^b_{0b}=E^b_{0c}=0.$
Setting $k'=d$ in (\ref{constraint1}) along with $k=c$ in (\ref{constraint2}) gives $B_c J_d = J_d B_c$, 
which in turn requires that
\begin{equation}
B_b = v J_b\ 
\label{BJ}
\end{equation}
for some constant $v$.  Relabeling $d\to c$ in (\ref{constraint2}) and comparing the two again requires the error term for non-vanishing $\epsilon_{kk'b}$ obey
\begin{equation}
 E^{b}_{kk'}  \widetilde{C}_{k'} C_k =  - E^{b}_{k'k}\widetilde{C}_k C_{k'} \ .
\label{ECC}
\end{equation}

To have non-vanishing coefficient of the commutator proportional to $\sigma^b\otimes \sigma^{c}$, either $a$ or $a'$ in the sum in $M_{k,k'}$ from \eqref{Mkkdef} must be 0. This coefficient is
\begin{align}
&\sum_{a',l}i \epsilon_{ba'c}J_b A^{0}_{kl}A^{a'}_{lk'}+ \sum_{a,l}i\epsilon_{cab}J_c A^{a}_{kl}A^0_{lk'}\cr
&\qquad=i\delta_{k0}\left(\epsilon_{bk'c}J_bC_{k'} + \epsilon_{ck'b}J_cC_{k'}B_{k'}\right)
+i \delta_{k'0}\big(
\epsilon_{bkc}J_bB_k \widetilde{C}_k+\epsilon_{ckb}J_c\widetilde{C}_k\big) \ .
\label{commrs}
\end{align}
The corresponding right-hand-side of (\ref{frank}) is
\begin{align}
\sum_m\big[E^{b}_{km} A^{c}_{mk'} - A^{b}_{km} E^{c}_{mk'}\big]=
\delta_{ck'}E^{b}_{k0} C_{c} + \delta_{k'0} E^{b}_{kc}\widetilde{C}_c
-\delta_{bk}\widetilde{C}_b E^c_{0k'} - \delta_{k0}C_b E^c_{bk'}
\label{RHSrs}
\end{align}
Since the commutator in \eqref{commrs} vanishes for $c=k'$ or $b=k$, and to satisfy these cases we need to impose $E^b_{00}=0$ in addition to the earlier conditions on the error terms. Imposing these conditions, the remaining terms in \eqref{commrs} and in \eqref{RHSrs} therefore have either a $\delta_{k0}$
or a $\delta_{k'0}$. Since \eqref{frank} must be true for any choice of $k,k'$, these two types of terms must each be equated, yielding 
\begin{align}
E^{b}_{dc}\widetilde{C}_c\ = i\epsilon_{bdc}\widetilde{C}_d\big(v J_b J_d-J_c\big)
\ ,\qquad -E^{c}_{bd}{C}_b\ = i\epsilon_{bdc}{C}_{d}\big(J_b- v J_{d} J_c\big)\ ,
\label{constraint3}
\end{align}
where we used \eqref{BJ}. We have relabelled $k=d$ in the former and $k'=d$ in the latter of these two constraints, as all the other cases are automatically satisfied because of our constraints on the error terms. Exchanging $b\leftrightarrow c$ in the latter shows the pair are consistent with (\ref{ECC}). Given \eqref{BJ}, we thus have two different independent relations for the error term $E^b_{dc}$ remaining, either \eqref{constraint1} or \eqref{constraint2}, and either of \eqref{constraint3}. Equating them gives 
\begin{align}
E^b_{dc} =  i\epsilon_{bdc}\frac{\widetilde{C}_d}{\widetilde{C}_c}\left(v J_b J_d-J_c\right)= 
 i\epsilon_{dcb}  \frac{\widetilde{C}_d C_c}{v}
\label{Ekpl}
\end{align}
We arrive at the final relation for the coefficients in the MPO,
\begin{equation}
{\widetilde{C}_c}C_c = vJ_c-v^2J_b J_d\ .
\label{CCu}
\end{equation}
with $b\ne c\ne d\ne b$. We thus have recovered the coefficients in \eqref{Acons}.

The last step is to check is $r=r'$ in \eqref{frank}, where the commutator vanishes. For $r=r'\ne 0$, the previous relations for the error term are sufficient to make the right-hand side vanish as well. For $r=r'=0$, setting $E^0_{kk'}=0$ as well suffices. Thus the only non-vanishing error terms come from \eqref{Ekpl}, yielding \eqref{Edef}. 

%We thus have a one-parameter family of MPOs obeying Frank's relation. These satisfy $A^0_{00}=1$ along with (\ref{BJ}) and (\ref{CCu}). The error term is given by (\ref{Ekpl}). There is a ``gauge'' choice, since (\ref{CCu}) leaves the ratio $C_{l}/\widetilde{C}_l$ undetermined. This choice does not affect the MPO if 00 or periodic boundary conditions are chosen. Otherwise, the choice is simply an overall constant. To recover the edge strong zero mode, we set $\widetilde{C}_z=0$ and impose $0z$ boundary conditions; the edge zero mode on the other end is found by setting $C_z=0$ and imposing $z0$ boundary conditions.

\section{Deriving the MPO for inhomogeneous couplings}
\label{app:impurity}

Here we give the details of the solution of \eqref{Sascha_Eq}, the consistency condition for the impurity Hamiltonian and the inhomogeneous MPO. We take the MPO to be of the form \eqref{Dform}, so that the non-vanishing coefficients of the Hamiltonian are those in \eqref{hform}.

We consider first the coefficients of $\sigma^0\sigma^b\sigma^0$ when $k_0=c$ and $k_3=d$. Because of \eqref{Edef}, the right-hand side vanishes, so that
\begin{align}
A^0_{dd}A^c_{c 0}D^d_{0d}h^{cc0}=A^0_{cc}A^d_{d 0}D^c_{c0}h^{0dd}\ .
\end{align}
These equations are solved by setting 
\begin{align}\label{Dbb0}
D^b_{b0}=h^{bb0}\frac{A^b_{b0}}{A^0_{bb}}\, \df \, \qquad\quad D^b_{0b}=h^{0bb}\frac{A^b_{0b}}{A^0_{bb}}\,\df\,,
\end{align}
for a function $\df$ yet to be determined.
We then consider the coefficients of $\sigma^b \sigma^c\sigma^0$ and $\sigma^0\sigma^b \sigma^c$, in both cases taking $k_0=b$ and $k_3=c$.  Using the above, we obtain the equations 
\begin{align}
D^{b}_{c d}A^0_{cc}A^0_{dd}= \df h^{c b d}A^{c}_{c0}A^{d}_{0d}
\end{align}
so that
\begin{align}
D^{b}_{c d}= h^{c b d}\frac{A^{c}_{c0}A^{d}_{0d}}{A^0_{cc}A^0_{dd}}f\,.
\end{align}

%\noindent\textbf{Step 2:} Next we consider the equation given by the indices $k_0=k_3\neq0 $ and the terms $\sigma^k\sigma^0\sigma^k$ together with $k_0=k_3\neq0 $ and $\sigma^k\sigma^\ell\sigma^m$ with $k\neq \ell$, $k\neq m$ $m\neq\ell$ after plugging in \eqref{fmskdnsjnd}. One can summarise the result of this by the equations with $k\neq\ell, k\neq m ,\ell\neq m$:
%\begin{align}
%h^{\ell \ell 0}h^{k \ell m}+h^{m \ell k}h^{0\ell \ell }=&0 \,,\qquad \qquad h^{\ell \ell 0}h^{k 0 k}-h^{m 0 m}h^{0\ell \ell }=0\,.\end{align}
%These twelve equations are solved by \begin{align}
%h^{kk0}=s^{(1)}_k\, h^{0kk}\,, \qquad  h^{k0k}=s^{(1)}_k\,Q \,, \qquad h^{\ell k m}=-s^{(1)}_k\,h^{mk\ell }\,,\qquad  s^{(1)}_k=\pm 1 \,,\qquad s^{(1)}_1s^{(1)}_2s^{(1)}_3=1\,. \end{align}

The next step is to analyse two cases in tandem. We consider the coefficients of $\sigma^b\sigma^{0}\sigma^b$ as well as those of $\sigma^b\sigma^c\sigma^d$.  Setting $k_0=k_3\ne 0$ and using \eqref{Dbb0}, we obtain
\begin{align}
h^{c c 0}h^{bcd}+h^{dcb}h^{0cc }=&0 \,,\qquad \qquad
h^{cc 0}h^{b 0 b}-h^{d 0 d}h^{0cc}=0\,.
\end{align}
These twelve equations are solved up to two arbitrary signs and a function $Q$:
\begin{align}
h^{bb0}=s_b\, h^{0bb}\,, \qquad  h^{b0b}=s_b\,Q \,, \qquad h^{c b d}=-s_b\,h^{dbc }\ ,
\label{signhere}
\end{align}
where  $s_k=\pm 1$ such that $s_xs_ys_z$\,=\,1.
We then use these results for the coefficients of $\sigma^b_1\sigma^0_2\sigma^0_3$ and $\sigma^0_1\sigma^0_2\sigma^b_3$ with $k_0=c$ and $k_3=d$, yielding
\begin{align}
D^{0}_{bb}A^0_{bb}=v\, s_b\, (Q\, D^0_{00}+\df\, (h^{bb0})^2) \,.
\end{align}
%We solve for
%\begin{align}
%D^{0}_{bb}=\frac{v\, s^{(1)}_b\, (Q\, D^0_{00}+\df\, (h^{bb0})^2)}{A^0_{bb}}
%\end{align}
We get further constraints on the impurity Hamiltonian by analysing three more cases. We consider the coefficients of $\sigma^c \sigma^0\sigma^d $ with $k_0=d$ and $k_3=c$, those of $\sigma^b \sigma^b\sigma^c$ with $k_0=d$ and $k_3=b$, and those of  $\sigma^c \sigma^b \sigma^b$ with $k_0=c$ and $k_3=d$. These cases all satisfy \eqref{Sascha_Eq} when
\begin{align}
h^{bb0}h^{c b d}=-h^{cc0}h^{ b c d}\,,\qquad\quad
Q h^{bb0}=-h^{c d b}h^{d c b}\ ,
\end{align}
both of which are solved by
 \begin{align}
h^{bb0}=-\frac{h^{c d b}h^{d c b}}{Q}\ .
\end{align}
%\noindent\textbf{Step 5:} We get further constraints on the Hamiltonian density by considering the equations $k_0=m$ and $k_3=\ell$ for $\sigma^\ell \sigma^0\sigma^m $ and $m\neq \ell$ and $k_0=m$ and $k_3=k$ with $\sigma^k \sigma^k\sigma^\ell $ or $\sigma^\ell \sigma^k \sigma^k$ and $k_0=\ell$ and $k_3=m$ \noindent\textbf{Step 6:} 
%Many combination now yield the same three equations of instances consider 

Finally, we consider $\sigma^0\sigma^b\sigma^b$ and $k_0=k_3=c$, yielding
\begin{align}\label{sdsndmsnd}
D^0_{00} A^0_{bb} Q v =A^b_{0b}A^{b}_{b0} \df Q +A^0_{bb}\df v (h^{c b d})^2\ .
\end{align}
This relation must hold for any $b$, so it results in three equations.  They are a little subtle to solve, as $h^{c b d}$ must be independent of $v$. Parametrizing $D^0_{00}=1-\alpha^2 v$, we find that $f=v$, and
%\begin{equation}
%\begin{aligned}
%D^{0}_{00}= d_0+\alpha^2 v+d_2 v^2+\dots\,,\\
%\df=f_0+f_1 v+f_2 v^2+\dots\,\,.
%\end{aligned}
%\end{equation}
%As we want to expand the equation in $v$, we need to insert the functional dependence of $A$ in $v$. Note that we do not need to pick a gauge as we have an explicit expression for $AA$. The zeroth order is trivially obeyed. The first order equation directly yields $f_0=0$. The second order of the equation gives us $f_1=d_0$. At the third order, we have multiple options. We can continue the series or have the option of truncating the series, i.e., stopping at the order $v$ in the power series of the variables, and use the remaining freedom in the Hamiltonian density to obey the equation of the order $v^3$. The higher orders are not generated. We proceed with the truncation and get the explicit expressions in terms of the homogeneous couplings $J_k$:
\begin{align}
\big(h^{c b d}\big)^2={\frac{Q}{J_b}}\big({J_c J_d-\alpha^2J_b}\big)\,
%,\qquad (c,b,d)=(2,1,3)\,,\ (1,3,2)\,,\ (3,2,1)\,. %\quad \text{and cyclic permutations}
\end{align}
where $\alpha$ must be independent of $v$. The relative sign under swap of the first and last index is determined from \eqref{signhere}. 
It is then straightforward to check that all remaining equations are solved by setting $Q=\alpha^2.$ Combining all these results gives the expressions in the main text.

\begin{comment}
We proceed with the truncation and get the explicit expressions in terms of the homogeneous couplings:
\begin{align}
h^{c k m}=\text{sig}(s^{(1)},s^{(2)_k})\frac{\sqrt{\alpha^2J_\ell J_m+f_1J_k}\sqrt{Q}}{\sqrt{\alpha^2}\sqrt{J_k}}\,
\end{align}
where the sig function is only a matter of sign factors: 
\begin{align}
\text{sig}(s^{(1)}_l,s^{(2)}_k)=((s^{(1)}_k+1)\epsilon^{\ell m}+(1-s^{(1)}_k))s^{(2)}_{k}
\end{align}
where $\epsilon^{12}=\epsilon^{13}=1$

%\noindent\textbf{Step 7:} All remaining equations are solved by the constraint 
%\begin{align}
%\alpha^2+d_0Q=0 \Longrightarrow Q=-\frac{\alpha^2}{d_0}
%\end{align}
\begin{align}
\alpha^2+d_0Q=0 \Longrightarrow Q=-\frac{\alpha^2}{d_0}
\end{align}
\noindent It turns out that the only variable which is important is the ratio $\frac{\alpha^2}{d_0}$ and so we set $d_0=1$ and $\alpha^2\to -\alpha^2$ to simplify the final expressions:   \\
\end{comment}

%%%%%%%%%%%%%%%%%%%%%%%%%%%%%%%%%%%%
\end{appendix}

\setlength{\bibsep}{3pt plus 0.3ex}

\bibliographystyle{JHEP}

\bibliography{XYZ}

\providecommand{\href}[2]{#2}\begingroup\raggedright\begin{thebibliography}{10}

\bibitem{Bethe1931}
H.~Bethe, \emph{{On the theory of metals. 1. Eigenvalues and eigenfunctions for
  the linear atomic chain}}, \href{https://doi.org/10.1007/BF01341708}{\emph{Z.
  Phys.} {\bfseries 71} (1931) 205}.

\bibitem{Orbach1958}
R.~Orbach, \emph{{Linear Antiferromagnetic Chain with Anisotropic Coupling}},
  \href{https://doi.org/10.1103/PhysRev.112.309}{\emph{Phys. Rev.} {\bfseries
  112} (1958) 309}.

\bibitem{Baxter1973a}
R.J.~Baxter, \emph{{Eight vertex model in lattice statistics and
  one-dimensional anisotropic Heisenberg chain. 1. Some fundamental
  eigenvectors}},
  \href{https://doi.org/10.1016/0003-4916(73)90439-9}{\emph{Annals Phys.}
  {\bfseries 76} (1973) 1}.

\bibitem{Baxter1973b}
R.J.~Baxter, \emph{{Eight vertex model in lattice statistics and
  one-dimensional anisotropic Heisenberg chain. 2. Equivalence to a generalized
  ice-type lattice model}},
  \href{https://doi.org/10.1016/0003-4916(73)90440-5}{\emph{Annals Phys.}
  {\bfseries 76} (1973) 25}.

\bibitem{Baxter1973c}
R.J.~Baxter, \emph{{Eight vertex model in lattice statistics and
  one-dimensional anisotropic Heisenberg chain. 3. Eigenvectors of the transfer
  matrix and Hamiltonian}},
  \href{https://doi.org/10.1016/0003-4916(73)90441-7}{\emph{Annals Phys.}
  {\bfseries 76} (1973) 48}.

\bibitem{Sutherland1970}
B.~Sutherland, \emph{Two‐dimensional hydrogen bonded crystals without the ice
  rule}, \href{https://doi.org/10.1063/1.1665111}{\emph{Journal of Mathematical
  Physics} {\bfseries 11} (1970) 3183}.

\bibitem{Verstraete2004}
F.~Verstraete, J.J.~Garc{\'\i}a-Ripoll and J.I.~Cirac, \emph{{Matrix Product
  Density Operators: Simulation of Finite-Temperature and Dissipative
  Systems}}, \href{https://doi.org/10.1103/PhysRevLett.93.207204}{\emph{Phys.
  Rev. Lett.} {\bfseries 93} (2004) 207204}
  [\href{https://arxiv.org/abs/cond-mat/0406426}{{\ttfamily
  cond-mat/0406426}}].

\bibitem{Pirvu_2010}
B.~Pirvu, V.~Murg, J.I.~Cirac and F.~Verstraete, \emph{Matrix product operator
  representations},
  \href{https://doi.org/10.1088/1367-2630/12/2/025012}{\emph{New Journal of
  Physics} {\bfseries 12} (2010) 025012}.

\bibitem{Cirac2020}
J.I.~Cirac, D.~Perez-Garcia, N.~Schuch and F.~Verstraete, \emph{{Matrix product
  states and projected entangled pair states: Concepts, symmetries, theorems}},
  \href{https://doi.org/10.1103/RevModPhys.93.045003}{\emph{Rev. Mod. Phys.}
  {\bfseries 93} (2021) 045003}
  [\href{https://arxiv.org/abs/2011.12127}{{\ttfamily 2011.12127}}].

\bibitem{Alicea2015}
J.~Alicea and P.~Fendley, \emph{{Topological phases with parafermions: theory
  and blueprints}},
  \href{https://doi.org/10.1146/annurev-conmatphys-031115-011336}{\emph{Ann.
  Rev. Condensed Matter Phys.} {\bfseries 7} (2016) 119}
  [\href{https://arxiv.org/abs/1504.02476}{{\ttfamily 1504.02476}}].

\bibitem{Fendley2015}
P.~Fendley, \emph{{Strong zero modes and eigenstate phase transitions in the
  XYZ/interacting Majorana chain}},
  \href{https://doi.org/10.1088/1751-8113/49/30/30LT01}{\emph{J. Phys. A}
  {\bfseries 49} (2016) 30LT01}
  [\href{https://arxiv.org/abs/1512.03441}{{\ttfamily 1512.03441}}].

\bibitem{Pfeuty1970}
P.~Pfeuty, \emph{{The one-dimensional Ising model with a transverse field}},
  \href{https://doi.org/10.1016/0003-4916(70)90270-8}{\emph{Annals Phys.}
  {\bfseries 57} (1970) 79}.

\bibitem{Kitaev2000}
A.~Kitaev, \emph{{Unpaired Majorana fermions in quantum wires}},
  \href{https://doi.org/10.1070/1063-7869/44/10S/S29}{\emph{Phys. Usp.}
  {\bfseries 44} (2001) 131}
  [\href{https://arxiv.org/abs/cond-mat/0010440}{{\ttfamily
  cond-mat/0010440}}].

\bibitem{Vernier2024}
E.~Vernier, H.-C.~Yeh, L.~Piroli and A.~Mitra, \emph{{Strong Zero Modes in
  Integrable Quantum Circuits}},
  \href{https://doi.org/10.1103/PhysRevLett.133.050606}{\emph{Phys. Rev. Lett.}
  {\bfseries 133} (2024) 050606}
  [\href{https://arxiv.org/abs/2401.12305}{{\ttfamily 2401.12305}}].

\bibitem{Katsura2015}
H.~Katsura, \emph{On integrable matrix product operators with bond dimension d
  = 4},
  \href{https://doi.org/10.1088/1742-5468/2015/01/P01006}{\emph{J.~Stat.~Mech.}
  {\bfseries 2015} (2015) P01006}
  [\href{https://arxiv.org/abs/1407.4262}{{\ttfamily 1407.4262}}].

\bibitem{Saleur1998}
H.~Saleur, \emph{{Lectures on nonperturbative field theory and quantum impurity
  problems}},  \href{https://arxiv.org/abs/cond-mat/9812110}{{\ttfamily
  cond-mat/9812110}}.

\bibitem{Baxter1982}
R.J.~Baxter, \emph{{Exactly solved models in statistical mechanics}} (1982),
  \href{https://doi.org/10.1142/9789814415255\_0002}{10.1142/9789814415255\_0002}.

\bibitem{Zhang:2022atz}
X.~Zhang, A.~Kl{\"u}mper and V.~Popkov, \emph{{Invariant subspaces and elliptic
  spin-helix states in the integrable open spin-1/2 XYZ chain}},
  \href{https://doi.org/10.1103/PhysRevB.106.075406}{\emph{Phys. Rev. B}
  {\bfseries 106} (2022) 075406}
  [\href{https://arxiv.org/abs/2204.05732}{{\ttfamily 2204.05732}}].

\bibitem{Zhang2023}
X.~Zhang, A.~Kl{\"u}mper and V.~Popkov, \emph{{Pedestrian's way to Baxter's
  Bethe ansatz for the periodic XYZ chain}},
  \href{https://doi.org/10.1103/PhysRevB.109.115411}{\emph{Phys. Rev. B}
  {\bfseries 109} (2024) 115411}
  [\href{https://arxiv.org/abs/2312.00161}{{\ttfamily 2312.00161}}].

\bibitem{essler2025strong}
F.H.L.~Essler, P.~Fendley and E.~Vernier, \emph{Strong zero modes in integrable
  spin-s chains with open boundaries},  2025.

\bibitem{Gehrmann2025}
S.~Gehrmann and F.H.L.~Essler, \emph{Exact strong zero modes in quantum
  circuits and spin chains with non-diagonal boundary conditions},  2025.

\bibitem{Corcoran2024}
L.~Corcoran, M.~de~Leeuw and B.~Pozsgay, \emph{{Integrable models on Rydberg
  atom chains}},
  \href{https://doi.org/10.21468/SciPostPhys.18.4.139}{\emph{SciPost Phys.}
  {\bfseries 18} (2025) 139}
  [\href{https://arxiv.org/abs/2405.15848}{{\ttfamily 2405.15848}}].

\bibitem{deLeeuw2024}
M.~de~Leeuw and V.~Posch, \emph{{All 4 $\times$ 4 solutions of the quantum
  Yang-Baxter equation}},  \href{https://arxiv.org/abs/2411.18685}{{\ttfamily
  2411.18685}}.

\bibitem{deLeeuw2020}
M.~de~Leeuw, C.~Paletta, A.~Pribytok, A.L.~Retore and P.~Ryan,
  \emph{{Yang-Baxter and the Boost: splitting the difference}},
  \href{https://doi.org/10.21468/SciPostPhys.11.3.069}{\emph{SciPost Phys.}
  {\bfseries 11} (2021) 069}
  [\href{https://arxiv.org/abs/2010.11231}{{\ttfamily 2010.11231}}].

\bibitem{Tetelman1982}
M.G.~Tetelman, \emph{Lorentz group for two-dimensional integrable lattice
  systems}, {\emph{Sov. Phys. JETP} {\bfseries 55} (1982) 306}.

\bibitem{10.1143/PTP.70.1508}
B.~Fuchssteiner, \emph{Master symmetries, higher order time-dependent
  symmetries and conserved densities of nonlinear evolution equations},
  \href{https://doi.org/10.1143/PTP.70.1508}{\emph{Progress of Theoretical
  Physics} {\bfseries 70} (1983) 1508}.

\bibitem{Links2001}
J.~Links, H.-Q.~Zhou, M.D.~Gould and R.H.~McKenzie, \emph{{Integrability and
  exact spectrum of a pairing model for nucleons}},
  \href{https://doi.org/10.1088/0305-4470/35/30/317}{\emph{J. Phys. A}
  {\bfseries 35} (2002) 6459}
  [\href{https://arxiv.org/abs/nlin/0110049}{{\ttfamily nlin/0110049}}].

\bibitem{Nienhuis2021}
B.~Nienhuis and O.E.~Huijgen, \emph{{The local conserved quantities of the
  closed XXZ chain}}, \href{https://doi.org/10.1088/1751-8121/ac0961}{\emph{J.
  Phys. A} {\bfseries 54} (2021) 304001}
  [\href{https://arxiv.org/abs/2104.01851}{{\ttfamily 2104.01851}}].

\bibitem{Nozawa2020}
Y.~Nozawa and K.~Fukai, \emph{{Explicit Construction of Local Conserved
  Quantities in the XYZ Spin-1/2 Chain}},
  \href{https://doi.org/10.1103/PhysRevLett.125.090602}{\emph{Phys. Rev. Lett.}
  {\bfseries 125} (2020) 090602}
  [\href{https://arxiv.org/abs/2003.02856}{{\ttfamily 2003.02856}}].

\bibitem{Fukai2025}
K.~Fukai and K.~Yamada, \emph{{to appear}},  2025.

\end{thebibliography}\endgroup

%\printbibliography
%\bibliographystyle{apsrev4-1}
%\bibliography{XYZ}
\end{document}